\documentclass[a4paper,11pt]{article}
\usepackage{pos}
\usepackage{hyperref}

\title{GRBAlpha, VZLUSAT-2 and GRBBeta - GRB observations with CubeSats}

\author*[a]{Jakub Řípa}
\author[a]{Marianna Dafčíková}
\author[b]{András Pál}
\author[a]{Norbert Werner}
\author[c]{Masanori Ohno}
\author[b]{László Mészáros}
\author[a]{Filip Münz}
\author[b]{Balázs Csák}
\author[g]{Gábor Galgóczi}
\author[a]{Nikola Husáriková}
\author[a]{Tomáš Vítek}
\author[a]{Pavel Kosík}
\author[a]{Michaela Ďuríšková}
\author[a]{Martin Kolář}
\author[a]{Lea Szakszonová}
\author[a,q]{Michal Pazderka}
\author[a]{Filip Hroch}
\author[s]{Martin Topinka}
\author[c]{Yasushi Fukazawa}
\author[c]{Hiromitsu Takahashi}
\author[c]{Tsunefumi Mizuno}
\author[c]{Masato Yokota}
\author[c]{Jean-Paul Breuer}
\author[o]{Kazuhiro Nakazawa}
\author[p]{Hirokazu Odaka}
\author[j]{Yuto Ichinohe}
\author[h]{Peter Hanák}
\author[r]{Miroslav Šmelko}
\author[u]{Ivo Veřtát}
\author[k]{Tomáš Urbanec}
\author[k]{Aleš Povalač}
\author[k]{Miroslav Kasal}
\author[f]{Jakub Kapuš}
\author[f]{Jan Hudec}
\author[f]{Marcel Frajt}
\author[f]{Maksim Rezenov}
\author[e]{Vladimír Dániel}
\author[e]{Petr Svoboda}
\author[e]{Juraj Dudáš}
\author[e]{Martin Sabol}
\author[l]{Róbert László}
\author[l]{Martin Koleda}
\author[d]{Hsiang-Kuang Chang}
\author[n]{Tsung-Che Liu}
\author[m]{Chih-Hsun Lin}
\author[i]{Chin-Ping Hu}
\author[t]{Che-Chih Tsao}
\author[d]{Kaustubha Sen}
\author[d]{Chih-En Wu}
\author[v]{Aaron Tohuvavohu}
\author[v]{Suresh Sivanandam}
\author[v]{Mark Barnet}

\affiliation[a]{Department of Theoretical Physics and Astrophysics, Faculty of Science, Masaryk University, Brno, Czech Republic}
\affiliation[b]{Konkoly Observatory, Research Centre for Astronomy and Earth Sciences, Budapest, Hungary}
\affiliation[c]{Department of Physics, Graduate School of Advanced Science and Engineering, Hiroshima University, Higashi-Hiroshima, Japan}
\affiliation[d]{Institute of Astronomy, National Tsing Hua University, Hsinchu, Taiwan, Republic of China}
\affiliation[e]{VZLU AEROSPACE, a.s., Prague, Czech Republic}
\affiliation[f]{Spacemanic Ltd., Bratislava, Slovakia}
\affiliation[g]{Wigner Research Centre for Physics, Budapest, Hungary}
\affiliation[h]{Faculty of Aeronautics, Technical University of Kosice, Košice, Slovakia}
\affiliation[i]{Department of Physics, National Changhua University of Education, Changhua City, Taiwan, Republic of China}
\affiliation[j]{RIKEN Nishina Center for Accelerator-Based Science, Saitama, Japan}
\affiliation[k]{Department of Radio Electronics, Faculty of Electrical Engineering and Communication, Brno University of Technology, Brno, Czech Republic}
\affiliation[l]{Needronix Ltd., Bratislava, Slovakia}
\affiliation[m]{Institute of Physics, Academia Sinica, Taipei, Taiwan, Republic of China}
\affiliation[n]{Institute of Physics, National Yang Ming Chiao Tung University, Hsinchu, Taiwan, Republic of China}
\affiliation[o]{Department of Physics, Nagoya University, Nagoya, Aichi, Japan}
\affiliation[p]{Department of Physics, The University of Tokyo, Bunkyo-ku, Tokyo, Japan}
\affiliation[q]{R\&D Center for Low-Cost Plasma and Nanotechnology Surface Modifications, CEPLANT, Department of Plasma Physics and Technology, Faculty of Science, Brno, Czech Republic}
\affiliation[r]{EDIS vvd., Košice, Slovakia}
\affiliation[s]{INAF - Osservatorio Astronomico di Cagliari, Selargius (CA), Italy}
\affiliation[t]{Department of Power Mechanical Engineering, National Tsing Hua University, Hsinchu, Taiwan, Republic of China}
\affiliation[u]{University of West Bohemia, Department of Applied Electronics and Telecommunications, Plzeň, Czech Republic}
\affiliation[v]{Dunlap Institute for Astronomy \& Astrophysics, University of Toronto, Toronto, Canada}

\emailAdd{ripa.jakub@gmail.com}

\abstract{
Results from GRBAlpha, VZLUSAT-2 and GRBBeta CubeSats and their on-board gamma-ray detectors for monitoring transients are summarised in this article. GRBAlpha was a 1U CubeSat launched in March 2021 to a 550 km altitude polar orbit carrying a CsI(Tl) scintillator gamma-ray burst (GRB) detector with a sensitive range of approximately 30-900 keV. It successfully operated for over four years until June 2025 when it de-orbited. VZLUSAT-2 was a 3U CubeSat launched in January 2022 to a 535 km altitude polar orbit and de-orbited in November 2025 after almost four years of smooth operation. It carried on board two GRB detectors very similar to the one used on GRBAlpha. Both missions have detected about 360 gamma-ray transients, including over 170 long and short gamma-ray bursts (GRBs), and including the most intense GRB ever recorded GRB 221009A and the second brightest GRB 230307A. The new family member, GRBBeta 2U CubeSat, integrated at Masaryk University, was launched in July 2024 to a 580 km altitude, 62 degree inclination orbit. It has been detecting GRBs since its launch without any trouble. Gamma-ray detectors on these nanosatellites are based on CsI(Tl) scintillator readout by silicon photomultipliers (SiPMs). These missions also provide a unique opportunity to study the radiation damage of SiPMs in the low Earth orbit environment and monitor the radiation belts. We have demonstrated that CubeSats can be used in missions lasting beyond three years and routinely detect GRBs.
}

\FullConference{Multifrequency Behaviour of High Energy Cosmic Sources - XV (MULTIF2025)\\
 9-14 June, 2025\\
Mondello, Palermo, Italy\\}

\begin{document}
\maketitle

\section{CubeSat Missions}\label{sec:missions}
The following sections describe GRBAlpha, VZLUSAT-2, and GRBBeta CubeSats carrying on board gamma-ray burst (GRB) detectors developed at Konkoly Observatory (Budapest, Hungary) with strong support from Hiroshima University (Higashi-Hiroshima, Japan), Masaryk University (Brno, Czech Rep.) and other partners from Japan. The GRB detectors on these missions are a technological verification of a novel scintillator-based detector read out by silicon photomultipliers (SiPMs). Each detector consists of a $75\times 75 \times 5$\,mm CsI(Tl) scintillator read out by eight multi-pixel photon counters (MPPCs) S13360-3050 PE by Hamamatsu Photonics K.K. The MPPCs comprise two independent readout channels with four MPPCs connected in parallel in each readout channel \cite{Pal2021, Pal2023, Ripa2025}. If these detectors are scaled up, they could be used on a proposed constellation of CubeSats called CAMELOT \cite{CAMELOT, Ohno2018}.

\subsection{GRBAlpha}

GRBAlpha\footnote{\href{https://grbalpha.konkoly.hu}{https://grbalpha.konkoly.hu}} \cite{Pal2021, Pal2023}, the smallest astrophysical space observatory, was a 1U CubeSat integrated at Konkoly Observatory (Budapest, Hungary) and launched in March 2021 to a 550 km altitude polar orbit. It carried one CsI(Tl) scintillator gamma-ray burst (GRB) detector with a sensitive range from $\sim 30$\,keV to $\sim 1$\,MeV. The satellite successfully operated for over four years until June 2025, when it naturally de-orbited.

\subsection{VZLUSAT-2}

VZLUSAT-2\footnote{\href{https://www.vzlusat2.cz}{https://www.vzlusat2.cz}} \cite{Daniel2020} was a 3U CubeSat developed by VZLU AEROSPACE, a.s., Prague, Czech Republic, with the purpose of verifying several technologies and payloads onboard. The satellite was launched in January 2022 to a 535 km altitude polar orbit and naturally de-orbited in November 2025 after almost four years of smooth operation. The telemetry and communication with the satellite were realised by a dedicated ground station of the Faculty of Electrical Engineering, University of West Bohemia \cite{Vertat2018}.

The satellite's primary payloads were two observing cameras. Secondary payloads were mainly composed of charged particle sensors, dosemeters and X-ray/gamma-ray sensors, e.g., a wide field-of-view X-ray/gamma-ray imager equipped with CdTe pixel semiconductor detector Timepix \cite{Granja2022}.

Among the secondary payloads were two GRB detectors placed under the solar panels. The GRB detectors were CsI(Tl) scintillators read out by MPPCs similar to the one used on GRBAlpha. The sensitivity range was from $\sim 30$\,keV to $\sim 2$\,MeV and in this work we report the transient detections by these CsI(Tl) based GRB detectors.

\subsection{GRBBeta}

The new family member, GRBBeta\footnote{\href{https://grbbeta.tuke.sk}{https://grbbeta.tuke.sk}} \cite{grbbeta} a 2U CubeSat, integrated at Masaryk University (Brno, Czech Rep.), was launched on July 9, 2024 to a 580\,km altitude, highly inclined orbit with $i=62^\circ$ on a maiden flight of Ariane-6. The primary goal of the GRBBeta mission is to continue detecting GRBs and additionally test new technological improvements crucial for future satellite constellations dedicated to GRB detection, such as S-band transmitter, the attitude determination and control system VAC by SERENUM Space, the Iridium satellite communication, and a system of four infrared sun sensors for precise attitude determination developed by the Konkoly Observatory.

The satellite carries on board one GRB detector very similar to the one on GRBAlpha with a sensitivity range from $\sim 50$\,keV to $\sim 2.3$\,MeV. The satellite has been detecting GRBs since its launch without any trouble.

The secondary scientific payload is a small space camera LUVCam \cite{Jeram2024} sensitive in the near ultraviolet range of $240-310$\,nm. It is a technology demonstrator and the in-orbit experiment to measure the behaviour of the CMOS Gpixel GSENSE4040 FSI imaging sensor in the Low Earth Orbit (LEO) environment. The same sensor, but backside-illuminated (BSI), is planned for use in the QUVIK ultraviolet space telescope mission \cite{Werner2024}. It was developed by the team from the Dunlap Institute for Astronomy \& Astrophysics at the University of Toronto.

\begin{figure}[h]
  \centering
  \includegraphics[height=4.9cm]{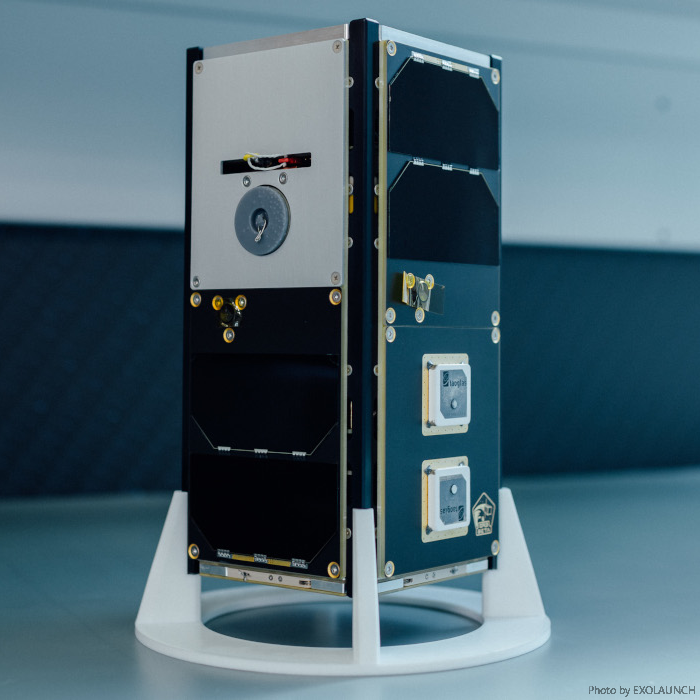}
  \includegraphics[height=4.9cm]{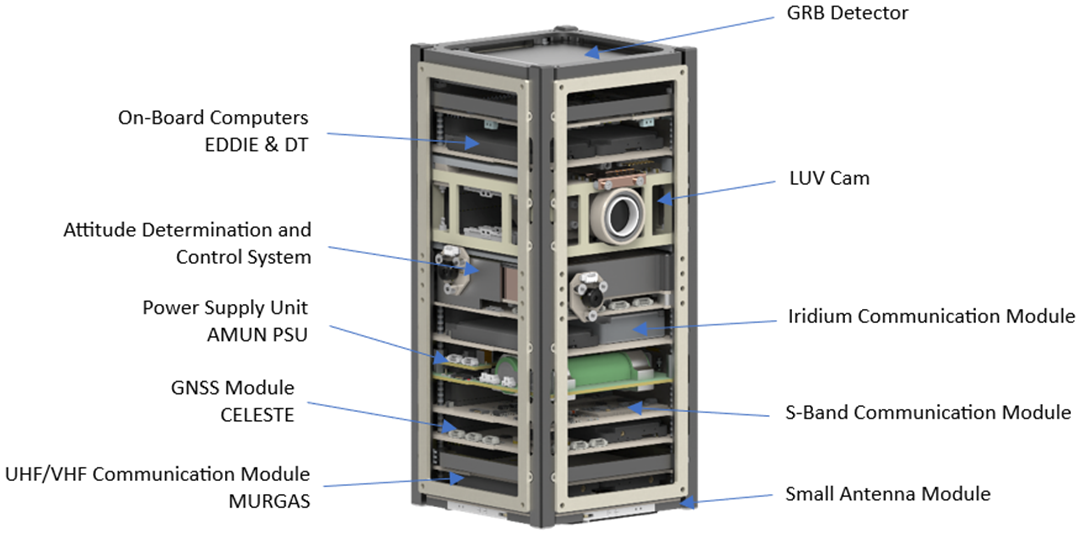}
  \caption{Left: A photo of the GRBBeta 2U CubeSat shortly before insertion into a satellite deployer (Credit: EXOLAUNCH). Right: 3D model of GRBBeta with description of the main components (Credit: SPACEMANIC).}
  \label{fig:GRBBeta}
\end{figure}

Figure~\ref{fig:GRBBeta} shows a photo of the integrated GRBBeta and 3D model with a description of the most important subsystems.

\section{Detected Gamma-Ray Transients}

While the ongoing results about the gamma-ray transients observed by GRBAlpha \& VZLUSAT-2 missions were reported in works \cite{Ripa2022, Munz2024}, the complete catalogue has been recently reported in work \cite{Dafcikova2026}. Table~\ref{tab:detections} presents the statistics of detected transients as of December 31, 2025. For GRBAlpha and VZLUSAT-2, these numbers can be considered final because both CubeSats have already de-orbited. In the case of GRBBeta, the numbers are current, as this mission continues to operate smoothly. Detected transients by these missions are also reported in the following websites: GRBAlpha\footnote{\href{https://monoceros.physics.muni.cz/hea/GRBAlpha/}{https://monoceros.physics.muni.cz/hea/GRBAlpha/}}, VZLUSAT-2\footnote{\href{https://monoceros.physics.muni.cz/hea/VZLUSAT-2/}{https://monoceros.physics.muni.cz/hea/VZLUSAT-2/}}, and GRBBeta\footnote{\href{https://monoceros.physics.muni.cz/hea/GRBBeta/}{https://monoceros.physics.muni.cz/hea/GRBBeta/}}.

\begin{table}
    \centering
    \begin{tabular}{cccc}
        Mission & GRBAlpha & VZLUSAT-2 & GRBBeta \\
        \hline
        \hline
        long GRBs    & 104 &  60 & 14 \\
        short GRBs   &  20 &   8 &  1 \\
        solar flares & 100 &  73 &  2 \\
        SGRs         &   2 &   5 &  0 \\
        X-ray binary &   1 &   0 &  0 \\
        \hline
        total        & 227 & 146 & 17 \\
    \end{tabular}
    \caption{Statistics of detected transients as of December 31, 2025. All detections include gamma-ray bursts (GRBs), solar flares, soft gamma repeaters (SGRs) and one X-ray binary outburst.}
    \label{tab:detections}
\end{table}

\begin{figure}[h]
  \centering
  \includegraphics[width=0.8\textwidth]{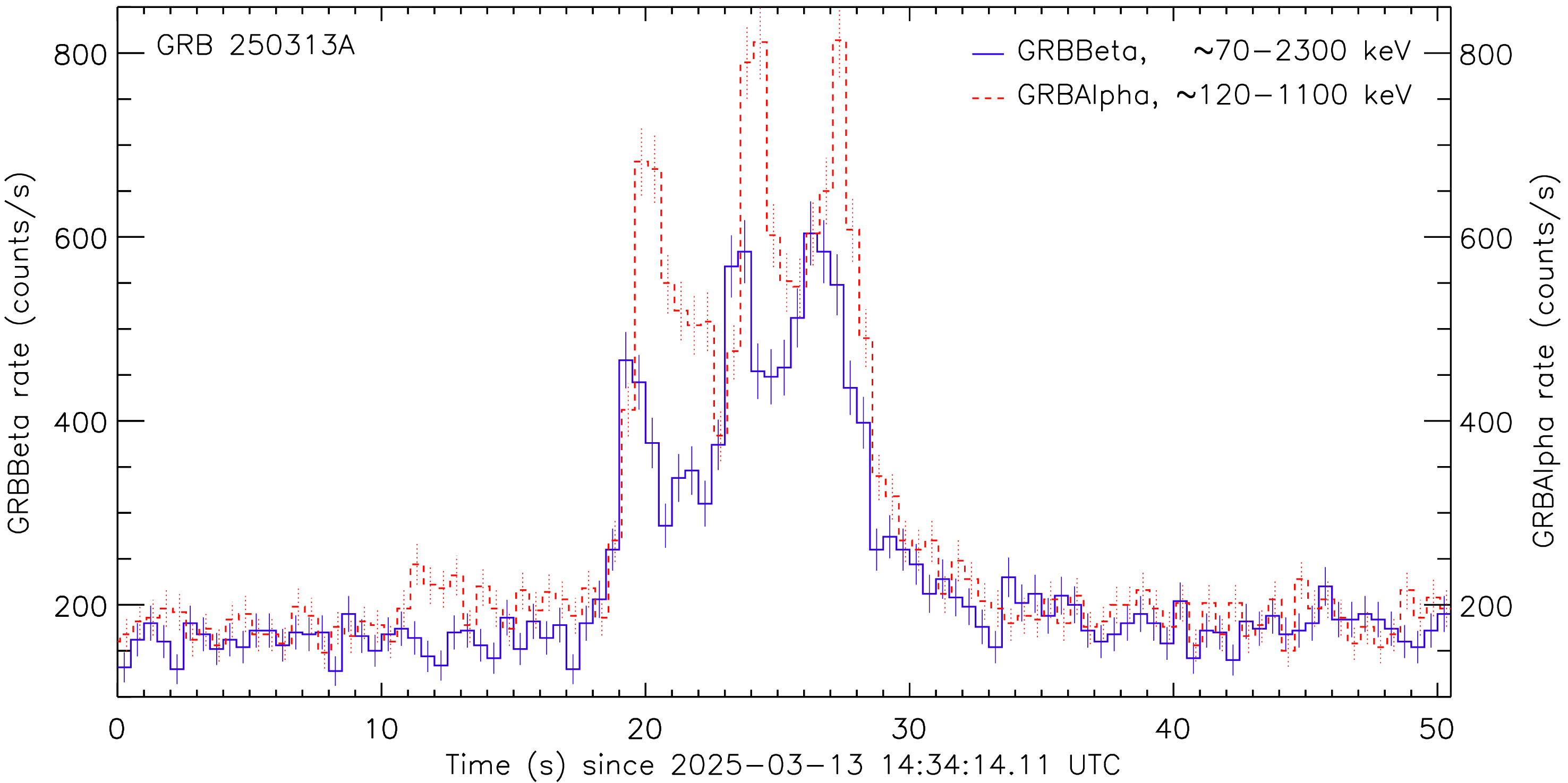}
  \caption{GRB 250313A coincidently detected by both CubeSats GRBAlpha and  GRBBeta.}
  \label{fig:GRB250313A}
\end{figure}

Twenty-eight gamma-ray transients were coincidently detected by GRBAlpha and VZLUSAT-2. One gamma-ray burst, GRB 250313A, was simultaneously observed by GRBAlpha and GRBBeta, see Figure~\ref{fig:GRB250313A}. Note that the shift of $\sim 0.5-1$\,s between the two light curves is probably due to imperfection in on-board clock synchronisation because the on-board clocks were not synchronised using the Global Positioning System (GPS), but by a command using an on-ground computer clock. Even this mini-constellation of two and later three CubeSats operating simultaneously has shown the potential of joint detections by multiple CubeSats if larger constellations of CubeSats are built in future. If the measurements of GRB light curves are obtained with high enough temporal resolution ($\sim 1$\,ms) and precise on-board clocks synchronisation, a triangulation method can be used for the localisation of GRB sources \cite{Ohno2018, Ohno2020, Sanna2021, Thomas2023}.

Among the exceptional GRB detections by these CubeSat missions are GRB 221009A and GRB 230307A.

GRB 221009A was the brightest burst ever recorded, nicknamed the ``Brightest Of All Time'' or the ``B.O.A.T.'' GRB \cite{Burns2023}. Remarkably, the Large High Altitude Air Shower Observatory (LHAASO) collaboration reported a detection of gamma rays up to 18\,TeV \cite{Huang2022}, in the following analysis the energy was lowered to 13\,TeV \cite{2023Sci...380.1390L, 2023SciA....9J2778C}, coming from the direction of the B.O.A.T. GRB. The GRBAlpha CubeSat detected this event too, and we were able to calculate the lower limit on the bolometric peak isotropic-equivalent luminosity in the $1-10,000$\,keV range (rest frame) to be $\geq8.4_{-1.5}^{+2.5}\times10^{52}$\,erg\,s$^{-1}$ (4\,s scale) and the lower limit on the isotropic-equivalent released energy to be $\geq2.8_{-0.5}^{+0.8}\times10^{54}$\,erg in the $1-10,000$\,keV band (rest frame) \cite{Ripa2023b}.

\begin{figure}[h]
  \centering
  \includegraphics[width=0.49\textwidth]{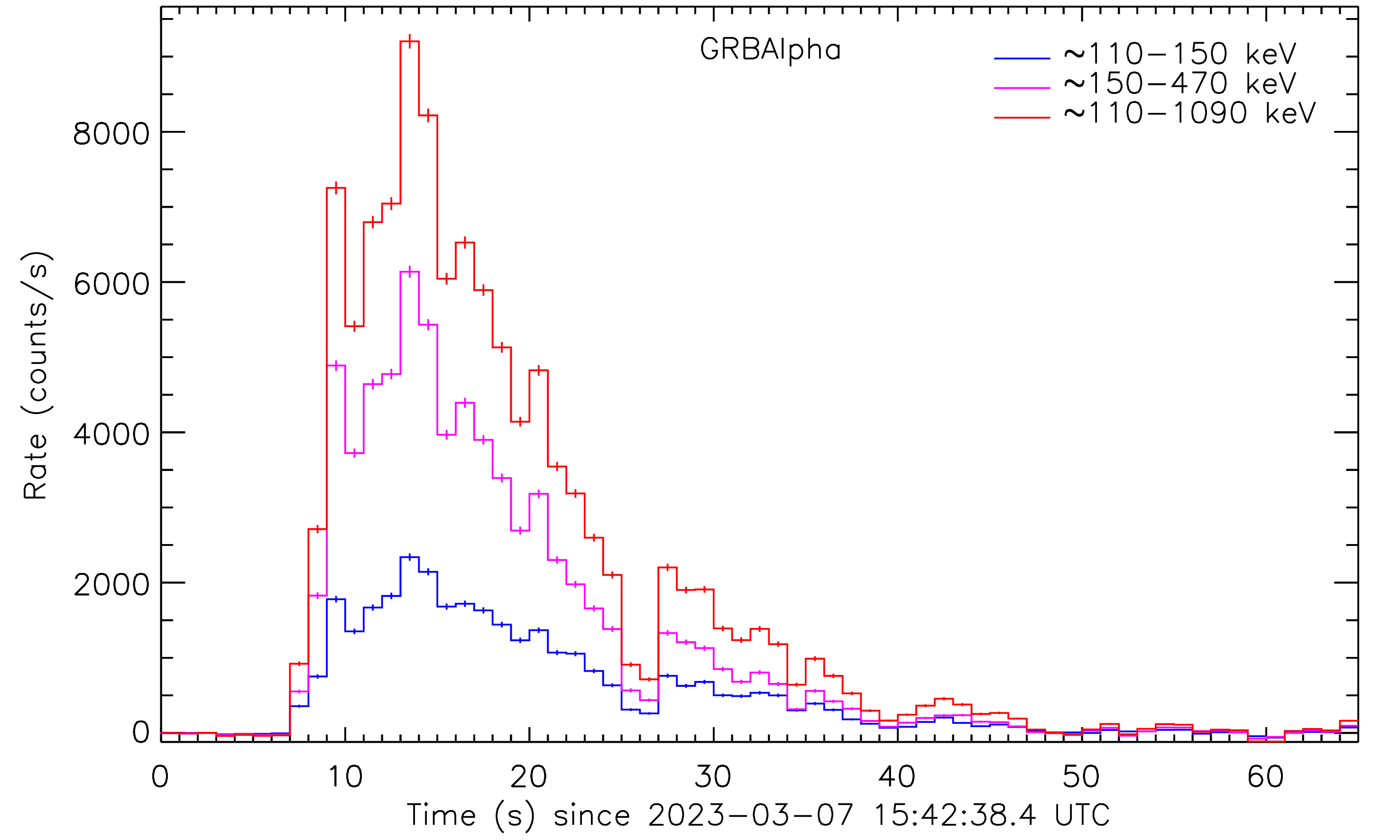}
  \includegraphics[width=0.49\textwidth]{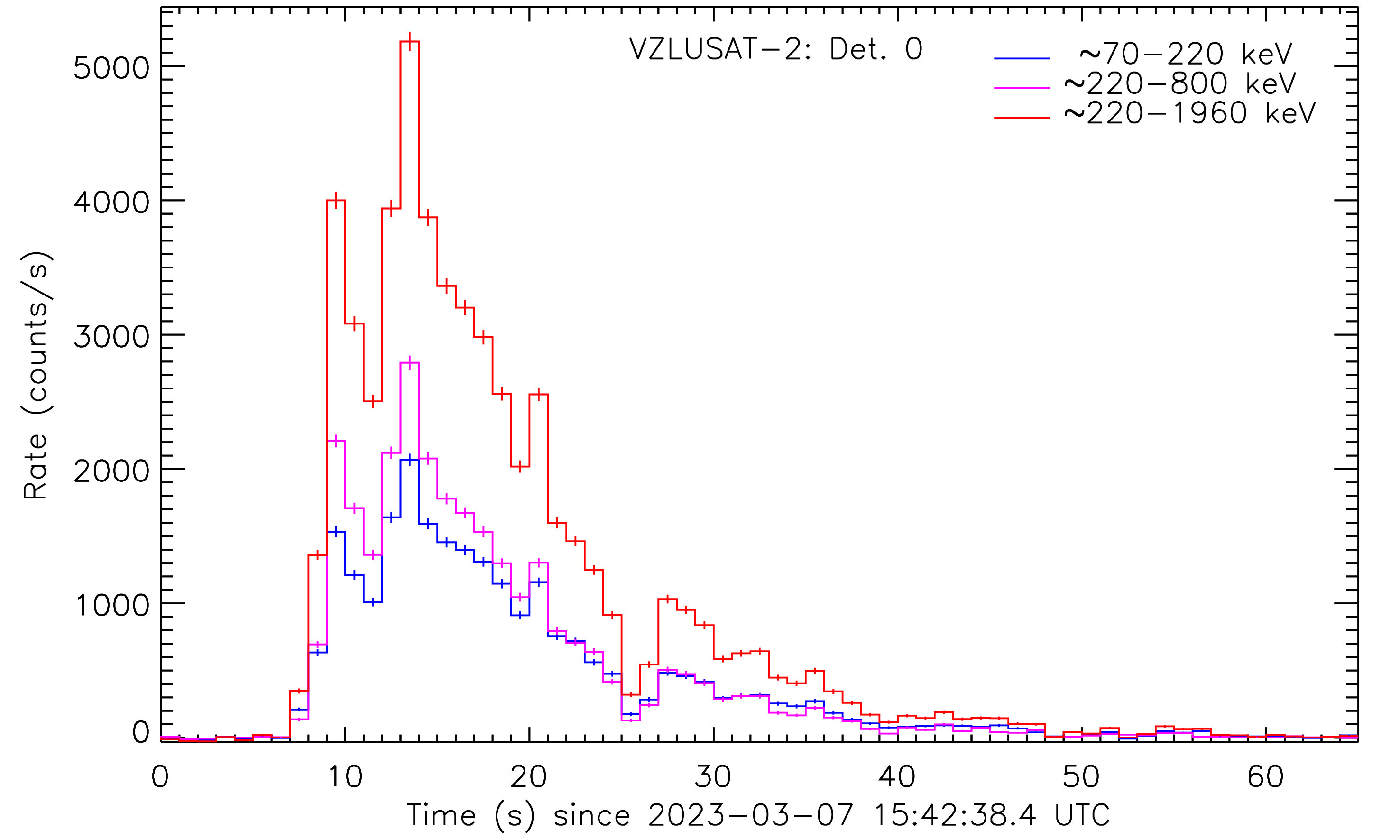}
  \caption{Light curves of the simultaneous detection of GRB 230307A by CubeSats GRBAlpha (left) and VZLUSAT-2 detector no. 0 (right).}
  \label{fig:GRB230307A}
\end{figure}

GRB 230307A was detected by both GRBAlpha and VZLUSAT-2 \cite{Dafcikova2023, Ripa2023a} and it was reported to be the second brightest GRB ever observed \cite{Dalessi2025}. Figure~\ref{fig:GRB230307A} shows the light curves of this GRB jointly observed by both missions. Remarkably, despite the long duration of the prompt gamma emission with $T_{90} \approx 35$\,s, this event has been associated with a kilonova detection pointing to a compact stellar merger origin \citep{Dai2024, Gillanders2025MNRAS, Levan2024Natur}.

\section{Radiation Environment at LEO and SiPM Degradation}

Besides the gamma-ray transient detections, these CubeSats allowed us to monitor the radiation background at LEO, reconstruct the radiation background maps, monitor the variation and enhancements of charged particle backgrounds at polar regions (outer Van Allen radiation belt) due to solar coronal mass ejections, as well as map the South Atlantic Anomaly (SAA). Preliminary radiation maps can be found in the work \cite{Munz2024}.

As mentioned in Sec .~\ref {sec:missions}, the GRB detectors on these CubeSats employ SiPMs to register the scintillation light produced in the CsI(Tl) crystals. SiPMs offer numerous advantages for usage in compact detectors, especially on CubeSats with limited space and power. However, it is known that they are sensitive to energetic protons, which causes degradation of their performance. Therefore, if used on a mission at LEO with high inclination, daily passages through SAA with high density of trapped energetic protons increase the dark current in SiPMs and leads to an increase in the low-energy sensitivity threshold. This has been studied in work \cite{Ripa2025}, where the degradation of the performance of Hamamatsu MPPCs S13360-3050 PE is characterised over more than three years in the polar low Earth orbit.

It should also be noted that the gain of MPPCs depends on temperature, which varies throughout the orbit and changes periodically between the sun-illuminated and dark parts of the orbit. In work \cite{Munz2024}, we describe the variation of the gain of the detector with respect to the temperature and operating voltage (bias voltage applied to MPPCs) studied on the flight spare GRBBeta detector by laboratory measurements.

\section{Conclusions}
\begin{itemize}
    \item We have demonstrated that CubeSats can be used in missions at LEO lasting beyond three years and routinely detect GRBs with the potential of simultaneous GRB detections by multiple satellites if larger constellations of CubeSats are built in future.
    \item We have flight-proven compact GRB detectors based on CsI(Tl) scintillator read out by SiPMs that can be readily implemented on CubeSat missions.
    \item We have demonstrated that SiPMs can be used in the LEO environment on a scientific mission lasting beyond three years when sufficient shielding is provided. This demonstrates the potential of SiPMs being used in future high-energy astrophysics space missions.
\end{itemize}

\acknowledgments
This work was supported by the Czech Science Foundation (GAČR) project No. 24-11487J. MD is a Brno Ph.D. Talent Scholarship Holder---Funded by the Brno City Municipality.


\bibliographystyle{JHEP}
\bibliography{references.bib}

\bigskip
\bigskip
\noindent {\bf DISCUSSION}

\bigskip
\noindent {\bf CARLOTTA PITTORI:} Do you have a publication regarding the background and radiation damage at LEO?

\bigskip
\noindent {\bf JAKUB ŘÍPA:} Our main publication about the degradation of SiPM at LEO is Řípa et al. 2025, Nuclear Instruments and Methods in Physics Research A, 1076, 170513 \cite{Ripa2025}. Our publications regarding the simulated background at LEO are Řípa et al. 2019, Astronomische Nachrichten, 340, 666 \cite{Ripa2019} and Galgóczi et al. 2021, Journal of Astronomical Telescopes, Instruments, and Systems, 7, id. 028004 \cite{Galgoczi2021}. The publications with preliminary measurements of background at LEO are Řípa et al. 2022, Proc. of SPIE, 12181, 121811K \cite{Ripa2022} and Münz et al. 2024, Proc. of SPIE, 13093, 130936J \cite{Munz2024}.


\bigskip
\noindent {\bf P. SAVINA:} What is the thickness of the SiPMs shielding?

\bigskip
\noindent {\bf JAKUB ŘÍPA:} The thickness of the shielding is 2-2.5 mm of PbSb3 alloy (depending on the side). It shields most of the trapped protons effectively, while cosmic rays are more problematic to shield.


\bigskip
\noindent {\bf SERGIO FABIANI:} How did you manage the variation of gain of SiPM with the variation of temperature?

\bigskip
\noindent {\bf JAKUB ŘÍPA:} We did not correct for temperature variation on board. In general, one can adjust the bias voltage or correct it on the ground by using temperature-dependent gain calibration data, provided the temperature is measured.

\end{document}